\documentclass{iopart}

\usepackage{amsfonts,amssymb}
\newenvironment{proof}{{\sl Proof.\,}}{\par\medskip }
\newtheorem{example}{Example}

\newtheorem{theorem}{{\bf Theorem.\,}}{\par\medskip }

\newtheorem{corollary}{{\bf Corollary.\,}}{\par\medskip }

\def\newpic#1{%
   \def\emline##1##2##3##4##5##6{%
      \put(##1,##2){\special{em:point #1##3}}%
      \put(##4,##5){\special{em:point #1##6}}%
      \special{em:line #1##3,#1##6}}}

\newpic{}

\def\bbbe{\Bbb E}
\def\bbbr{{\Bbb R}}

\def\bbbc{{\Bbb C}}
\def\bbbe{{\Bbb E}}
\def\bbbz{{\Bbb Z}}
\def\bbbd{{\Bbb D}}
\def\wedgecomma{\mathop{\wedge}\limits_{'}}
\def\tr{\mbox{tr}\,}
\def\re{{\rm Re}\,}
\def\im{{\rm Im}\,}
\def\Ad{\mbox{Ad}\,}
\def\Aut{\mbox{Aut}\,}
\def\diag{\mbox{diag}\,}

\def\fr#1{{\mathfrak{#1}}}
\def\openone{\leavevmode\hbox{\small1\kern-3.3pt\normalsize1}}

\eqnobysec

\begin{document}

\title[$N $-wave interactions and simple Lie
algebras]{ $N $-wave interactions related to simple Lie
algebras.\\
$\bbbz_2$- reductions and soliton solutions}

\author{V. S. Gerdjikov$^\dag$, G. G. Grahovski$^\dag$, R. I.
Ivanov$^\ddag$\footnote[3]{On leave of absence
from Institute for Nuclear Research and Nuclear Energy, Bulgarian Academy
of Sciences, 72 Tsarigradsko chaussee, 1784 Sofia, Bulgaria } and N. A.
Kostov$^ \#$ }


\address{\dag\  Institute for Nuclear Research and Nuclear Energy,
\\ Bulgarian Academy of Sciences, 72 Tsarigradsko chaussee, 1784 Sofia,
Bulgaria }

\address{\ddag \ Department of Mathematical Physics, \\ National
University of Ireland-Galway, Galway, Ireland}

\address{\# \ Institute of Electronics, Bulgarian Academy of Sciences,\\
72 Tsarigradsko chaussee, 1784 Sofia, Bulgaria}

\begin{abstract}

The reductions of the integrable $N $-wave type equations solvable by the
inverse scattering method with the generalized Zakharov-Shabat systems $L
$ and related to some simple Lie algebra $\fr{g} $ are analyzed.
The Zakharov-Shabat dressing method is extended to the case when $\fr{g} $
is an orthogonal algebra. Several types of one  soliton solutions of the
corresponding $N$-wave equations and their reductions are studied.  We
show that to each soliton solution one can relate a (semi-)simple
subalgebra of $\fr{g} $.  We illustrate our results by $4 $-wave
equations related to $so(5) $ which find applications in
Stockes-anti-Stockes wave generation.

\end{abstract}
\jl{5}


\maketitle

\section{Introduction}

One of the important nonlinear models with numerous applications to
physics that appeared at the early stages of development of the inverse
scattering method (ISM), see \cite{ZM,ZM1,K,1,FaTa,KRB} is the
3-wave resonant interaction model described by the equations:
\begin{eqnarray}\label{eq:*1}
{\partial u_1 \over \partial t} + v_1 {\partial u_1 \over \partial x} =
i\varepsilon \bar{u}_2 u_3(x,t), \nonumber\\
{\partial u_2 \over \partial t} + v_2 {\partial u_2 \over \partial x} =
i\varepsilon \bar{u}_1 u_3(x,t), \\
{\partial u_3 \over \partial t} + v_3 {\partial u_3 \over \partial x} =
i\varepsilon u_1 u_2(x,t). \nonumber
\end{eqnarray}
Here $v_i $ are the group velocities and $\varepsilon  $ is the
interaction constant.

The 3-wave equations can be solved through the ISM due to the fact that
Eq.~(\ref{eq:*1}) allows Lax representation (see eq.~(\ref{eq:LM}) below).
The main result of the pioneer papers \cite{ZM,ZM1} consist also in
proving that if $u_k(x,t) $, $k=1,2,3 $ satisfy the system (\ref{eq:*1})
then the one-parameter family of ordinary differential operators $L_1(t)$:
\begin{eqnarray}\label{eq:*2}
L_1(t)\psi (x,t,\lambda ) &\equiv  \left( i {d \over d x}
+q(x,t) - \lambda J \right) \psi (x,t,\lambda )= 0, \\
q(x,t) &= \left( \begin{array}{ccc} 0 & q_{12} & q_{13} \\
q_{12}^* & 0 & q_{23} \\ q_{13}^* & q_{23}^* & 0 \end{array} \right),
\qquad J = \left( \begin{array}{ccc} J_1 & 0 & 0 \\ 0 & J_2 & 0 \\ 0 &
0 & J_3 \end{array} \right), \nonumber
\end{eqnarray}
are isospectral. Here $J_1>J_2>J_3 $, $q_{12}(x,t) =
u_1(x,t) \sqrt{J_1-J_2}$, $q_{13}(x,t) = u_3(x,t)\sqrt{J_1-J_3} $ and
$q_{23}(x,t) = u_2(x,t)\sqrt{J_2-J_3} $. If we denote by $T_1(\lambda ,t)
$ the transfer matrix of $L_1(t) $ then $T_1(\lambda ,t) $ evolves in time
according to the linear equation:
\begin{equation}\label{eq:*3}
i {dT_1  \over dt } - \lambda [I,T_1(\lambda ,t)] = 0, \qquad
I = \left( \begin{array}{ccc} I_1 & 0 & 0 \\ 0 & I_2 & 0 \\ 0 &
0 & I_3 \end{array} \right).
\end{equation}
The group velocities $v_k$ are expressed through the constants $J_k $ and
$I_k $ by $v_1=-v_{12} $, $v_2=-v_{23} $ and $v_3=-v_{13} $ where $v_{jk}
= (I_j-I_k)/(J_j-J_k) $. Thus the problem of solving the $3 $-wave
equation (\ref{eq:*1}) for a given initial condition $u_k(x,t=0) =
u_{k,0}(x) $ can be performed in three steps:

\begin{description}

\item {\bf a)} Insert $u_{k,0}(x) $ as potential coefficients in $L_1(0) $
and solve the scattering problem for $L_1(0) $ determine the initial value
of transfer matrix $T_1(\lambda ,0) $;

\item {\bf b)} Solve (\ref{eq:*3}) and determine $T_1(\lambda ,t) $ for $
t>0 $;

\item {\bf c)} Solve the inverse scattering problem for $L_1(t) $, i.e.
reconstruct the  potential $q(x,t) $ corresponding to $T_1(\lambda ,t)
$ and, recover the solution $u_k(x,t) $ of (\ref{eq:*1}).

\end{description}

Note that step b) is solved trivially; steps a) and c) reduce to
linear problems. Thus the nonlinear three-wave interaction model is
reduced to a sequence of linear problems. In step a) one
must solve the direct scattering problem for $L_1(0) $ while in step c)
one should solve the inverse scattering problem for the operator $L_1(t)$.
Step c) for the operator $L_1(t) $ in the class of potentials vanishing
fast enough for $x\to \pm\infty  $ was first solved by Zakharov and
Manakov \cite{ZM1} by deriving the analog of the
Gel'fand-Levitan-Marchenko equation for $L_1(t)$. This was rather tedious
procedure but soon Shabat \cite{Sh} proved that step c) can be reduced to
a local Riemann-Hilbert problem (RHP) for the fundamental analytic
solutions of $L_1(t) $. This fact was used in \cite{Za*Sh} to simplify
greatly the derivation of the soliton solutions of the 3-wave system by
reducing it to a simple purely algebraic procedure known now as the
Zakharov-Shabat dressing method.

Quite naturally the 3-wave interaction model was generalized to
$N $--wave equations which can be written in the form:
\begin{eqnarray}\label{eq:1.4}
i[J,Q_t] - i[I,Q_x] +[[I,Q],[J,Q]] = 0,
\end{eqnarray}
where $Q(x,t)=-Q^\dag (x,t) $ is an off-diagonal $n\times n $
matrix-valued function (i.e. $Q_{jj}=0 $) tending fast enough to 0 for
$x\to \pm \infty $. The potential $q(x,t) $ in $L $ is replaced by
$[J,Q(x,t)] $ and $I $ and $J $ are constant diagonal matrices:
\begin{equation}\label{eq:IJ}
\fl I = \diag (I_1, I_2,\dots , I_n), \qquad
J = \diag (J_1, J_2,\dots , J_n), \qquad J_1>J_2>\dots >J_n,
\end{equation}
satisfying $\tr I = \tr J =0 $. In today's literature the system
(\ref{eq:1.4}) is known as the $N$-wave system with $N=n(n-1)/2 $. From
algebraic point of view the $3 $-wave system can be related to the algebra
$sl(3) $ while the $N $-wave system is related to $sl(n) $.

Due to the comparatively simple structure of the underlying algebra
$sl(n)$ it was rather straightforward to generalize not only the ISM but
also the RHP approach and the Zakharov-Shabat dressing method
\cite{1,Sh,Za*Sh}. Soon after that in \cite{G*86} it was proved that the
transition from the potential $q(x,t) $ to the corresponding transfer
matrix $T(\lambda ,t) $ can be viewed as a generalized Fourier transform
which allows one to analyze the hierarchy of Hamiltonian structures of the
$N $-wave systems. The next stage, namely the proof of the complete
integrability of these systems and the derivation of their action-angle
variables was done first in \cite{ZM} and later by different method in
\cite{BS}.

The next generalization, also rather natural from algebraic point of
view, restricts $Q(x,t) $, $I $  and $J $ in (\ref{eq:1.4}) to be elements
of a (semi-)simple Lie algebra $\fr{g} $; then $N=|\Delta_+ | $ -- the
number of positive roots of $\fr{g} $.  Though algebraically simple, this
restriction makes both the construction of the fundamental analytic
solution (FAS) and the dressing method more difficult, see \cite{VG*87}
where some preliminary results on the form of $u(x,\lambda ) $ are
reported. The difficulties are due to the fact that both methods require
explicit construction and/or factorizing of certain group elements which
even for the symplectic and orthogonal algebras is not trivial.

We should mention also the papers by Zakharov and Mikhailov \cite{Za*Mi}
in which they generalized the dressing method and derived the soliton
solutions for a number of field theory models related to the orthogonal
and symplectic algebras. Their analysis is based on  Lax operators more
complicated than (\ref{eq:1.1}) and needs to be modified in order to apply
it to $L $ (\ref{eq:1.1}).

The $N $-wave equation (\ref{eq:1.4}) related to $\fr{g} $ are solvable by
the ISM \cite{1} applied to the  generalized system of Zakharov--Shabat
type:
\begin{eqnarray}\label{eq:1.1}
\fl L(\lambda )\Psi(x,t,\lambda ) = \left(i{d \over dx} + [J,Q(x,t)]
- \lambda J \right)\Psi(x,t,\lambda ) = 0, \quad J\in \fr{h},\\
\label{eq:1.3.1}
\fl q(x,t)\equiv [J,Q(x,t)] = \sum_{\alpha \in \Delta _+} (\alpha ,\vec{J})
(Q_{\alpha }(x,t)E_{\alpha } - Q_{-\alpha}(x,t) E_{-\alpha }) \in \fr{g}
\backslash \fr{h},
\end{eqnarray}
where $ \fr{h}$ is the Cartan subalgebra and $E_{\alpha } $ are the root
vectors of the simple Lie algebra $\fr{g} $. Here and below $r =
\mbox{rank}\,\fr{g} $, $\Delta _+ $ is the set of positive roots of
$\fr{g} $ and $\vec{J}=\sum_{k=1}^{r} J_k e_k$, $ \vec{I}=\sum_{k=1}^{r}
I_k e_k  \in \bbbe^r$ are vectors corresponding to the Cartan elements
$J, I \in \fr{h}$.

The results in \cite{G*86} were generalized to any semi-simple Lie algebra
in \cite{VG*86}. The important question addressed and answered there
concerned the construction of the FAS for each of the fundamental
representations of $\fr{g} $. To this end one needs the explicit formulae
for the Gauss decompositions of the scattering matrix $T(\lambda ,t) $
which in turn requires knowledge of the fundamental representations of
$\fr{g} $. The next step that could be done with the FAS is the explicit
construction of the resolvent of $L(t) $ (see \cite{LMP}) and the
generalized Fourier expansions \cite{VG*86} which underlie all basic
properties of the nonlinear evolution equations (NLEE).

However all results in \cite{VG*86} were derived under the assumption
that $L(t) $ has no discrete eigenvalues. The problem is to find
a correct ansatz for the Zakharov-Shabat dressing factor $u(x,\lambda ) $
whose form depends on the choice of the representation of $\fr{g}$.

The numbers $|\Delta_+ | $ for the simple Lie algebras given in the
table below
\[ \arraycolsep=6pt
\begin{array}{|c|c|c|c|c|c|c|c|c|}\hline
\fr{g} & {\bf A}_r & {\bf B}_r, {\bf C}_r &  {\bf D}_r & {\bf G}_2 &
{\bf F}_4 &  {\bf E}_6 & {\bf E}_7 & {\bf E}_8 \\ \hline
|\Delta_+ | & r(r+1)/2 & r^2 & r(r-1) & 6 & 24 & 36 & 63 & 120 \\ \hline
\end{array}
\]
grow rather quickly with the rank of the algebra $r $. If we disregard the
constraint $Q^\dag =-Q $ the corresponding generic NLEE are systems of
$2|\Delta_+| $ equations for $2|\Delta_+| $ independent complex-valued
functions. They are solvable for any $r $ but their possible applications
to physics for large $r $ do not seem realistic. However one still may
extract new integrable and physically useful NLEE by imposing reductions
on $L(t) $,  i.e.  algebraic restrictions on $Q(x,t) $ which diminish the
number of independent functions in them  and the number of equations
\cite{2}. Of course such restrictions must  be compatible with the
dynamics of the NLEE.  One of the simplest and best known reductions
that we already mentioned $Q^\dag =-Q $ diminishes the number of
fields by a factor of 2.  Another famous class of reductions found long
ago \cite{2,MiOlPer} led to the integrability of the 2-dimensional Toda
chains. The related group of reductions is isomorphic to $\bbbd_h $ where
$h $ is the Coxeter number of the corresponding Lie algebra $\fr{g} $. As
a result the number of independent real-valued fields becomes equal to the
rank of the algebra.

Although the two reductions outlined above have been known for quite a
time, comparatively little is known about the other `intermediate' types
of reductions.  One of the aims of this paper is to outline how this
gap could be bridged. The ingredients that we need are the well known
facts about the simple Lie algebras and their representations.  Another
aim is to extend the Zakharov-Shabat dressing method for linear systems
related to orthogonal algebras. Several types of one soliton solutions of
the corresponding $N$-wave equations and their reductions are analyzed.

Section~2 contains preliminaries from the ISM, the
reduction group \cite{2}, the theory of simple Lie algebras and the
scattering theory for $L $.  In Section~3 we extend the Zakharov-Shabat
dressing method for linear systems (\ref{eq:1.1}) related to
the orthogonal algebras. We provide the general form of the
one-soliton solutions of the corresponding $N $-wave equations.  In
Section~4 we formulate the effect of $G_R $ on the scattering data of the
Lax operator and analyze two types of $\bbbz_2 $-reductions.  In Section~5
we analyze the consequences of the $\bbbz_2 $-reductions on the soliton
parameters and exhibit several types of 1-soliton solutions related to
subalgebras $sl(2) $, $so(3)$ and $sl(3) $ of $\fr{g} $.
The $4 $-wave equations related to $so(5) $ are shown to have
applications to physics. In the last
Section~6 we briefly discuss the hierarchies of the Hamiltonian
structures.


\section{Preliminaries and general approach}\label{2}

\subsection{Lax representations and reductions}\label{ssec:2.1}

Indeed the $N $-wave equations (\ref{eq:1.4}) as well as the other
members of the hierarchy possess Lax representation of the form:
\begin{equation}\label{eq:LM}
[L(\lambda ),M_P(\lambda )] = 0,
\end{equation}
where $L(\lambda ) $ is provided by (\ref{eq:1.4}) and
\begin{eqnarray}\label{eq:1.3}
\fl M_P(\lambda )\Psi(x,t,\lambda ) \equiv \left( i{d \over dt} +
\sum_{k=0}^{P-1}V_k(x,t) - f_P\lambda^P I\right) \Psi(x,t,\lambda ) = 0,
\quad I \in \fr{h} .
\end{eqnarray}
Eq.~(\ref{eq:LM}) must hold identically with respect to $\lambda  $. A
standard procedure generalizing the AKNS one \cite{AKNS} allows us to
evaluate $V_k $ in terms of $Q(x,t) $ and its $x $-derivatives. Here and
below we consider only the class of smooth potentials $Q(x,t) $ vanishing
fast enough for $|x|\to \infty $ for any fixed value of $t $. Then one may
also check that the asymptotic value of the potential in $M_P $, namely
$f^{(P)}(\lambda ) =f_P\lambda ^PI $ may be understood as the dispersion
law of the corresponding NLEE. The $N $-wave equations (\ref{eq:1.4}) is
obtained in the simplest nontrivial case with $P=1 $, $f_P=1 $ and
$V_0(x,t)=[I,Q(x,t)]$.

The consistent approach to the reduction problem is based on the
notion of the reduction group $G_R $ introduced in \cite{2} and further
developed in \cite{ForGib,Za*Mi,ForKu}. Since we impose
finite number of algebraic constraints on the potentials $U $ and $V_P $
\begin{equation}\label{eq:4.3}
\fl U(x,t,\lambda )= [J,Q(x,t)] - \lambda J, \qquad  V_P(x,t,\lambda ) =
\sum_{k=0}^{P-1} V_k(x,t) \lambda ^k - f_P \lambda ^P I,
\end{equation}
of the Lax pair it is natural to choose as $G_R $ finite group
which must allow for two realizations: i)~as finite subgroup of
the group $\Aut(\fr{g}) $  of automorphisms of the algebra
$\fr{g} $; and ii)~as finite subgroup of the conformal mappings
$\mbox{Conf\,} \bbbc$ on the complex $\lambda $-plane. Obviously
to each reduction imposed on $L $ and $M $ there corresponds a
reduction of the space of fundamental solutions $\fr{S}_\Psi
\equiv \{\Psi (x,t,\lambda )\} $ to (\ref{eq:1.1}) and
(\ref{eq:1.3}). Some of the simplest $\bbbz_2 $-reductions
(involutions) of $N$-wave systems have been known for a long time
(see \cite{2}) and are related to outer automorphisms of $\fr{g}
$ and $\fr{G} $, namely:
\begin{equation}\label{eq:C-1}
\fl C_1\left( \Psi (x,t,\lambda ) \right) =  A_1 \Psi^\dag (x,t,\kappa_1
(\lambda )) A_1^{-1} = \tilde{\Psi}^{-1}(x,t,\lambda ),
\qquad \kappa _1(\lambda )=\pm\lambda ^*,
\end{equation}
where $A_1 $ belongs to the Cartan subgroup of the group $\fr{G} $:
\begin{equation}\label{eq:A-1}
A_1 = \exp \left( \pi i H_1 \right).
\end{equation}
Here $H_1 \in \fr{h}$ is such that $\alpha (H_1)\in \bbbz $ for all roots
$\alpha \in \Delta $ in the root system  $\Delta  $ of ${\frak g} $.
Note that the reduction condition relates one fundamental solution
$\Psi(x,t,\lambda )\in \fr{G} $ of (\ref{eq:1.1}) and (\ref{eq:1.3})  to
another $\tilde{\Psi}(x,t,\lambda ) $ which in general differs from
$\Psi(x,t,\lambda ) $.

Another class of $\bbbz_2 $ reductions are related to automorphisms of the
type:
\begin{equation}\label{eq:C_2}
\fl C_2\left( \Psi (x,t,\kappa _2(\lambda) ) \right) \equiv A_2 \Psi^T
(x,t,\kappa_2 (\lambda )) A_2^{-1} = \tilde{\Psi}^{-1}(x,t,\lambda ),
\qquad \kappa _2(\lambda )=\pm\lambda ,
\end{equation}
where $A_2 $ may be of the form (\ref{eq:A-1}) or corresponds to a Weyl
group automorphism.  The best known examples of NLEE obtained with the
reduction (\ref{eq:C_2}) are the sine-Gordon and the MKdV equations which
are related to $\fr{g}\simeq sl(2) $. For higher rank algebras such
reductions to our knowledge have not been studied.

Since our aim is to preserve the form of the Lax pair we consider only
automorphisms preserving the Cartan subalgebra ${\frak  h} $. This
condition is obviously fulfilled if: a)~$ A_k $, $k=1,2 $ is in the form
(\ref{eq:A-1}); b)~$A_k $, $k=1,2 $ are Weyl group automorphisms. Most of
our results below concern the orthogonal algebras which we define by
$X+SX^TS^{-1} =0 $ with
\begin{eqnarray}\label{eq:br-dr}
\fl && S=\sum_{k=1}^{r} (-1)^{k+1} (E_{k\bar{k}} + E_{\bar{k}k}) +
(-1)^{r}E_{r+1,r+1}, \nonumber\\
\fl && \qquad \bar{k}= N+1 -k , \qquad N=2r+1 \qquad \fr{g}\simeq {\bf
B}_r, \\
\fl && S=\sum_{k=1}^{r} (-1)^{k+1} (E_{k\bar{k}} + E_{\bar{k}k})
, \qquad N=2r, \qquad \bar{k}= N+1 -k , \qquad \fr{g}\simeq {\bf D}_r,
\nonumber
\end{eqnarray}
Here $E_{kn} $ is an $N\times N $ matrix whose
matrix elements are $(E_{kn})_{ij}=\delta _{ik}\delta _{nj} $ and $N $ is
the dimension of the typical representation of the corresponding algebra.


The corresponding reduction conditions for the Lax pair
are as follows \cite{2}:
\begin{equation}\label{eq:2.1}
C_k(U(\Gamma _k(\lambda ))) = U(\lambda ), \qquad
C_k(V_P(\Gamma _k(\lambda ))) = V_P(\lambda ),
\end{equation}
where $C_k\in \mbox{Aut}\; \fr{g} $ and $\Gamma _k(\lambda )$ are
the images of $g_k $.
Since $G_R $ is a finite group then for each $g_k $ there exist an
integer $N_k $ such that $g_k^{N_k} =\openone $; if $k=1 $ and $N_1=2 $
then $G_R\simeq \bbbz_2 $.

\subsection{Cartan-Weyl basis and Weyl group}

Here we fix up the notations and the normalization conditions for the
Cartan-Weyl generators of $\fr{g} $. The commutation relations are
given by \cite{LA,Helg}:
\begin{eqnarray}\label{eq:31.2}
&& [h_k,E_\alpha ] = (\alpha ,e_k) E_\alpha , \quad [E_\alpha ,E_{-\alpha
}]=H_\alpha , \nonumber\\
&& [E_\alpha ,E_\beta ] = \left\{ \begin{array}{ll}
N_{\alpha ,\beta } E_{\alpha +\beta } \quad & \mbox{for}\; \alpha +\beta
\in \Delta \\ 0 & \mbox{for}\; \alpha +\beta \not\in \Delta
\cup\{0\}. \end{array}  \right.
\end{eqnarray}

If $J$ is a regular real element in $\fr{h} $ then we may  use it to
introduce an ordering in $\Delta  $ by saying that the root $\alpha
\in\Delta _+ $ is positive (negative) if $(\alpha ,\vec{J})>0 $ ($(\alpha
,\vec{J})<0 $ respectively).  The normalization of the basis is determined
by:
\begin{eqnarray}\label{eq:32.1}
&& E_{-\alpha } =E_\alpha ^T, \quad \langle E_{-\alpha },E_\alpha \rangle
={2  \over (\alpha ,\alpha ) }, \nonumber\\
&& N_{-\alpha ,-\beta } = -N_{\alpha ,\beta }, \quad N_{\alpha ,\beta } =
\pm (p+1),
\end{eqnarray}
where the integer $p\geq 0 $ is such that $\alpha +s\beta \in\Delta  $ for
all $s=1,\dots,p $ and $ \alpha +(p+1)\beta \not\in\Delta  $.
The root system $\Delta  $ of $\fr{g} $ is invariant with respect to the
Weyl reflections $S_\alpha  $; on the vectors $\vec{y}\in \bbbe^r $ they
act as
\begin{equation}\label{eq:32.2}
S_\alpha \vec{y} = \vec{y} - {2(\alpha ,\vec{y})  \over (\alpha ,\alpha )}
\alpha , \quad \alpha \in \Delta .
\end{equation}
All Weyl reflections $S_\alpha  $ form a finite group $W_{\fr{g}} $ known
as the Weyl group. The Weyl group has a natural action of the on the
Cartan-Weyl basis:
\begin{eqnarray}\label{eq:32.3}
\fl S_\alpha (H_\beta ) \equiv A_\alpha H_\beta A^{-1}_{\alpha } =
H_{S_\alpha \beta }, \qquad
S_\alpha (E_\beta ) = n_{\alpha ,\beta } E_{S_\alpha \beta }, \quad
n_{\alpha ,\beta }=\pm 1.
\end{eqnarray}
which shows how it can be understood as a group of inner automorphisms of
$\fr{g} $ preserving the Cartan subalgebra $\fr{h} $.
The same property is possessed also by $\Ad_{\fr{h}} $ automorphisms;
indeed, choosing $K=\exp(\pi i H_{\vec{c}})$ and from (\ref{eq:31.2}) one
finds
\begin{equation}\label{eq:35.2}
KH_\alpha K^{-1} = H_\alpha , \quad
KE_\alpha K^{-1} = e^{\pi i (\alpha ,\vec{c})} E_\alpha ,
\end{equation}
where $\vec{c}\in \bbbe^r $ is the vector corresponding to
$H_{\vec{c}} \in \fr{h}$. Then the condition $K^2=\openone  $
means that $(\alpha ,\vec{c})\in \bbbz $ for all $\alpha \in
\Delta  $. If $\omega _k $ are the fundamental weights of $\fr{g}
$ then $H_{\vec{c}}$ must be such that $\vec{c}=\sum_{k=1}^{r}
2c_k\omega _k/(\alpha _k,\alpha _k) $ with $c_k $ integer.

We will need also the element $w_0\in W(\fr{g}) $,
$w_0^2=\openone  $ which maps the highest weight of each irreducible
representation $\Gamma (\omega ) $ to the corresponding lowest weight,
i.e.,
\begin{eqnarray}\label{eq:w0.1}
\fl w_0(E_{\alpha }) = n_{\alpha }E_{w_0(\alpha) }, \quad w_0(H_k)=
H_{w_0(e_k)}, \quad \alpha  \in \Delta _+, \quad n_{\alpha }= \pm 1;
\end{eqnarray}
The action of $w_0 $ on Cartan-Weyl basis is given by (\ref{eq:32.3}) with
$S_\alpha  $ replaces by $S $ (\ref{eq:br-dr}). On the root space $w_0 $
is defined by $w_0(e_k)=-e_k $ for all $k=1,\dots, r $ if $\fr{g}\simeq
{\bf B}_r $ or ${\bf D}_{2r} $; for ${\bf D} _{2r+1}$ we have
$w_0(e_k)=-e_k $ for $k=1,\dots,r-1 $ and $w_0(e_r)=e_r $, see
\cite{LA,Helg}.

\subsection{The inverse scattering problem for $L $}

The direct scattering problem for the Lax operator (\ref{eq:1.1}) is
based on the Jost solutions
\begin{equation}\label{eq:5.1}
\lim_{x\to\infty } \psi (x,\lambda )e^{i\lambda Jx} = \openone ,\qquad
\lim_{x\to-\infty } \phi (x,\lambda )e^{i\lambda Jx} = \openone ,
\end{equation}
and the scattering matrix
\begin{equation}\label{eq:5.2}
T(\lambda )=(\psi (x,\lambda ))^{-1} \phi (x,\lambda ).
\end{equation}
The FAS $\chi ^\pm(x,\lambda )$ of $L(t) $ are analytic functions of
$\lambda $ for $\im \lambda \gtrless 0 $ are related to the Jost solutions
by:
\begin{equation}\label{eq:5.6}
\chi ^\pm(x,\lambda ) =\phi (x,\lambda ) S^\pm (\lambda ) =
\psi (x,\lambda ) T^\mp (\lambda ) D^\pm(\lambda )
\end{equation}
where by "hat" above we denote the inverse matrix $\hat{T}\equiv T^{-1} $.
Here $S^\pm(\lambda ) $, $D^\pm(\lambda ) $ and $T^\pm(\lambda ) $ are
the factors in the Gauss decomposition of the scattering matrix
\begin{eqnarray}\label{eq:5.3}
\fl && T(\lambda )= T^-(\lambda ) D^+(\lambda ) \hat{S}^+(\lambda )
= T^+(\lambda ) D^-(\lambda ) \hat{S}^-(\lambda ),\\
\label{eq:5.4}
\fl && S^\pm (\lambda ) = \exp \left( \sum_{\alpha \in\Delta _+}^{}
s_{\alpha }^{\pm}(\lambda ) E_{\pm\alpha }\right) , \quad
T^\pm (\lambda ) = \exp \left( \sum_{\alpha \in\Delta _+}^{}
t_{\alpha }^{\pm}(\lambda ) E_{\pm\alpha }\right) , \\
\label{eq:5.5}
\fl && D^+ (\lambda ) = \exp \left( \sum_{j=1}^{r}
{2d_j^{+}(\lambda ) \over (\alpha _j,\alpha _j)} H_{j}\right) ,
\quad D^- (\lambda ) = \exp \left( \sum_{j=1}^{r}
{2d_j^{-}(\lambda ) \over (\alpha _j,\alpha _j)}H_{j}^-\right) ,
\end{eqnarray}
where $H_j \equiv H_{\alpha _j}$, $ H_j^- = w_0(H_j) $. The proof
of the analyticity of $\chi ^\pm(x,\lambda ) $ for any
semi-simple Lie algebra and real $J $ is given in \cite{VG*86}.
The superscripts $+ $ and $- $ in $D^\pm (\lambda ) $ shows that
$D_j^+(\lambda) $ and $D_j^-(\lambda) $:
\begin{equation}\label{eq:2.13'}
D_j^\pm (\lambda ) = \langle \omega _j^\pm |D^\pm (\lambda )|\omega^\pm
_j\rangle = \exp \left( d_j^\pm (\lambda )\right), \qquad
\omega ^-_j = w_0(\omega ^+_j),
\end{equation}
are analytic functions of $\lambda $ for $\im \lambda >0 $ and $\im
\lambda <0 $ respectively. Here $\omega ^+_j $ are the fundamental weights
of $\fr{g} $ and $|\omega _j^+\rangle $ and $|\omega _j^-\rangle $ are the
highest and lowest weight vectors in these representations.
On the real axis $\chi ^+(x,\lambda ) $ and $\chi ^-(x,\lambda ) $ are
related by
\begin{equation}\label{eq:5.7}
\chi ^+(x,\lambda )= \chi ^-(x,\lambda ) G_0(\lambda ), \qquad
G_0(\lambda ) =S^+(\lambda ) \hat{S}^-(\lambda ),
\end{equation}
and the sewing function $G_0(\lambda ) $ may be considered as a minimal
set of scattering data provided the Lax operator (\ref{eq:1.1}) has no
discrete eigenvalues. The presence of discrete eigenvalues $\lambda _1^\pm
$ means that one (or more) of the functions $D_j^\pm(\lambda ) $
will have zeroes at $\lambda _1^\pm $, for more details see
\cite{G*86}. If we introduce $\xi^\pm(x,\lambda )=\chi ^\pm(x,\lambda )
e^{i\lambda Jx} $ then eq.~(\ref{eq:5.7}) can be cast in the form:
\begin{equation}\label{eq:5.8}
\fl \xi^+(x,\lambda )= \xi^-(x,\lambda ) G(x,\lambda ), \qquad
G(x,\lambda ) = e^{-i\lambda Jx} G_0(\lambda ) e^{i\lambda Jx},
\qquad \lambda \in \bbbr.
\end{equation}
This relation together with the normalization condition
\begin{equation}\label{eq:5.8a}
\lim_{\lambda \to\infty } \xi^\pm(x,\lambda ) = \openone ,
\end{equation}
can be interpreted as a RHP with canonical normalization

If the potential $q(x,t) $ of the Lax operator (\ref{eq:1.1}) satisfies
the $N $-wave equation (\ref{eq:1.4}) then $S^\pm(t,\lambda ) $,
$T^\pm(t,\lambda ) $ satisfy the linear evolution equations (\ref{eq:*3}),
while the functions $D^\pm (\lambda ) $ are time-independent. Therefore
$D_j^\pm(\lambda ) $ can be considered as the generating functions of the
integrals of motion of (\ref{eq:1.4}).

\subsection{Hamiltonian properties of the NLEE}

The interpretation of the ISM as a generalized Fourier transform and the
expansions over the ``squared solutions'' of (\ref{eq:1.1}) were derived
in \cite{G*86}. All $N $-wave type equations are Hamiltonian and possess
a hierarchy of pair-wise compatible Hamiltonian structures $\{H^{(k)},
\Omega ^{(k)}\} $, $k=0,\pm 1,\pm 2, \dots$.  The phase space
${\cal  M} $ of these equations is the space spanned by the
complex-valued functions $\{ Q_\alpha , \alpha \in \Delta \}$,
$\dim_\bbbc {\cal  M}=|\Delta | $. The corresponding NLEE as, e.g.
(\ref{eq:1.4}) and its higher analogs can be formally written down as
Hamiltonian equations of motion:
\begin{equation}\label{eq:cH}
\Omega ^{(k)} (Q_t,\cdot ) = dH^{(k)} (\cdot), \qquad k=0, \pm1, \pm 2,
\dots,
\end{equation}
where both $\Omega ^{(k)} $ and $H^{(k)} $ are complex-valued. The
simplest Hamiltonian formulation of (\ref{eq:1.4}) is given by
$\{H^{(0)} $, $\Omega^{(0)}\}$ where $H^{(0)}=H_0 + H_{\rm int} $ and
\begin{eqnarray}\label{eq:1.5}
\fl H_0 ={c_0 \over 2i}\int_{-\infty }^{\infty } \, dx\, \left\langle
Q,[I,Q_x] \right\rangle = ic_0 \int_{-\infty }^{\infty }dx\,\sum_{\alpha
>0}^{ }\, {(\vec{b},\alpha )\over (\alpha ,\alpha )} (Q_{\alpha }
Q_{-\alpha ,x} - Q_{\alpha ,x}Q_{-\alpha }), \\
\label{eq:H-om}
\fl H_{\rm int} = {c_0\over 3} \int_{-\infty }^{\infty }\, dx \,
\left\langle [J,Q],[Q,[I,Q]] \right\rangle = \sum_{[\alpha ,\beta, \gamma
]\in {\cal M}}\omega _{\beta ,\gamma } H(\alpha ,\beta ,\gamma) ;\\
\fl H(\alpha ,\beta ,\gamma ) = c_0 \int_{-\infty}^{\infty } dx\,
(Q_{\alpha }Q_{-\beta }Q_{-\gamma } - Q_{-\alpha }Q_{\beta }Q_{\gamma }),
\quad \omega _{\beta \gamma } ={4N_{\beta ,\gamma }\over (\alpha
,\alpha)}\, \mbox{det}\left|\begin{array}{cc} (\vec{a},\beta ) &
(\vec{b},\beta  ) \\ (\vec{a},\gamma  ) & (\vec{b},\gamma )\end{array}
\right|, \nonumber
\end{eqnarray}
and the symplectic form $\Omega ^{(0)}$ is equivalent to a canonical one
\begin{eqnarray}\label{eq:Ome}
\Omega^{(0)} &=& {ic_0 \over 2} \int_{-\infty }^{\infty } dx\, \left\langle
[J, \delta Q(x,t)] \wedgecomma \delta Q(x,t) \right\rangle = \nonumber\\
&=&i \sum_{\alpha \in \Delta _+} {2c_0(\vec{a},\alpha ) \over
(\alpha ,\alpha )}
\int_{-\infty }^{\infty }dx\, \delta Q_{\alpha }(x,t) \wedge
\delta Q_{-\alpha }(x,t).
\end{eqnarray}
Here $c_0 $ is a constant adjusted so that both $H^{(0)} $ and $\Omega
^{(0)} $ be real, $\langle \, \cdot\, , \cdot \, \rangle $ is the Killing
form of $\fr{g}$ and the triple $[\alpha ,\beta ,\gamma ] $ belongs to
${\cal M} $ if $\alpha ,\beta ,\gamma \in \Delta _+ $ and $\alpha =\beta
+\gamma $.

For the $N $-wave equations and their higher analogs $H^{(k)} $ depend
analytically on $Q_\alpha $. That allows one to rewrite the
equation (\ref{eq:cH}) as a standard Hamiltonian equation with real-valued
$\Omega ^{(k)} $ and $H^{(k)} $. The phase space then is viewed by the
manifold of real-valued functions $\{ \re Q_\alpha ,\im Q_\alpha \}$,
$\alpha \in \Delta  $, so $\dim _\bbbr {\cal  M}=2|\Delta | $. Such
treatment is formal and we will not explain it into more details here.

Another well known way to make $\Omega ^{(k)} $ and $H^{(k)} $ real is
to impose reduction on them involving complex or hermitian conjugation as
in (\ref{eq:C-1}).

Physically to each term $H(\alpha ,\beta ,\gamma ) $ we relate part of a
wave-decay diagram which shows how the wave associated with the root
$\alpha $ decays into $\beta  $ and $\gamma  $ waves.  We assign to
each root $\alpha  $ an wave with wave number $k_{\alpha } $ and frequency
$\omega _{\alpha } $. Each of the elementary decays preserves them, i.e.
\[ k_{\alpha } = k_{\beta } + k_{\gamma }, \qquad \omega (k_{\alpha}) =
\omega (k_{\beta }) + \omega (k_{\gamma }).
\]

Thus the number of the different wave types and their decay modes are
determined by the properties of the system of positive roots $\Delta _+ $
of $\fr{g} $.

\section{The dressing Zakharov-Shabat method}

The main goal of the dressing method is, starting from a FAS
$\chi ^\pm_0(x,\lambda ) $ of $L $ with potential $q_{(0)}=[J,Q_{(0)}] $ to
construct a new singular solution $\chi ^\pm_1(x,\lambda )$ of the RHP
(\ref{eq:5.8}), (\ref{eq:5.8a}) with singularities located at prescribed
positions $\lambda _1^\pm $. The new solutions $\chi ^\pm_1(x,\lambda )$
will correspond to a potential $q_{(1)}=[J,Q_{(1)}] $ of $L $
(\ref{eq:1.1}) with two discrete eigenvalues $\lambda _1^\pm $.  It is
related to the regular one by a dressing factor $u(x,\lambda )$
\begin{eqnarray}\label{eq:Dressfactor}
\chi^{\pm}_1(x,\lambda)=u(x,\lambda) \chi^{\pm}_0(x,\lambda)
u_{-}^{-1}(\lambda ), \qquad u_{-}(\lambda )=\lim_{x\to -\infty }
u(x,\lambda ).
\end{eqnarray}
The sewing function in the RHP is modified to $G_1(x,\lambda ) =
u_-(\lambda )G(x,\lambda )u^{-1}(\lambda ) $; as we will see below
$u_-(\lambda ) $ is an element of the Cartan subgroup of $\fr{G} $.
Then $u(x,\lambda ) $ obviously must satisfy the equation
\begin{eqnarray}\label{eq:u-eq}
i {du  \over dx } + q_{(1)}(x) u(x,\lambda ) - u(x,\lambda ) q_{(0)}(x) -
\lambda [J,u(x,\lambda )] =0,
\end{eqnarray}
and the normalization condition $\lim_{\lambda \to\infty } u(x,\lambda )
=\openone $. Besides $\chi ^\pm_i (x,\lambda ) $, $i=0,1 $ and
$u(x,\lambda ) $ must belong to the corresponding group ${\frak  G} $; in
addition $u(x,\lambda ) $ by construction has poles and/or zeroes at
$\lambda_1^\pm$.  Below all quantities related to $L $ (\ref{eq:1.1}) with
potential $q_i(x) $ will be supplied by the corresponding index $i $.
Their scattering data are related by:
\begin{eqnarray}\label{eq:SD1-0}
\fl && S_{(1)}^{\pm}(\lambda ) = u_-(\lambda ) S_{(0)}^{\pm} (\lambda )
u_-^{-1}(\lambda ), \qquad T_{(1)}^{\pm} (\lambda ) = u_+(\lambda )
T_{(0)}^{\pm}(\lambda ) u_+^{-1}(\lambda ), \nonumber\\
\fl && D_{(1)}^{\pm}(\lambda ) = u_+(\lambda ) D_{(0)}^{\pm} (\lambda )
u_-^{-1}(\lambda ), \qquad u_{\pm}(\lambda ) = \lim_{x\to \pm\infty }
u(x,\lambda ) .
\end{eqnarray}
Since the limits $u_\pm(\lambda ) $ are $x $-independent and belong to
the Cartan subgroup of $\fr{G} $, so $S_{(1)}^{\pm}(\lambda )  $,
$T_{(1)}^{\pm}(\lambda )$ are of the form (\ref{eq:5.4}), (\ref{eq:5.5}).

The construction of $u(x,\lambda ) $ will be based on an appropriate
ansatz specifying explicitly the form of its $\lambda  $-dependence which
crucially depends on the choice of $\fr{g} $ and its representation. Here
we will consider separately the classical series  of simple Lie algebras:
${\bf A}_r \simeq sl(r+1) $, ${\bf B}_r \simeq so(2r+1) $ and ${\bf D}_r
\simeq so(2r) $. The simplest nontrivial case $\fr{g}\simeq {\bf A}_r  $
is solved in the classical papers \cite{Za*Sh,1} with the ansatz
\begin{eqnarray}\label{eq:sl-n}
u(x,\lambda )= \openone + (c_1(\lambda )-1) P_1(x), \qquad
c_1(\lambda ) = {\lambda -\lambda _1^+  \over \lambda -\lambda _1^- },
\end{eqnarray}
where the rank 1 projector $P_1(x) $ is of the form:
\begin{eqnarray}\label{eq:pr-1}
&& P_1(x)= {|n(x)\rangle \langle m(x)|\over \langle m(x)|n(x)\rangle },
\nonumber\\
&& |n(x)\rangle =\chi _0^+(x,\lambda_1^+)|n_{0}\rangle, \qquad
\langle m(x)|= \langle m_{0}| \hat{\chi} _0^-(x,\lambda _1^-).
\end{eqnarray}
where $|n_0\rangle  $ and $\langle m_0| $ are constant vectors.
It can be easily checked that $u(x,\lambda ) $ (\ref{eq:sl-n}) corresponds
to a potential
\begin{eqnarray}\label{eq:q-1}
q_{(1)}(x) = q_{(0)}(x) + \lim_{\lambda \to \infty} (J -u(x,\lambda ) J
u^{-1}(x,\lambda )).
\end{eqnarray}

In fact $u(x,\lambda ) $ (\ref{eq:sl-n}) belongs not to $SL(r+1) $, but to
$GL(r+1) $.  It is not a problem to multiply $u(x,\lambda ) $ by an
appropriate scalar and thus to adjust its determinant to 1. Such a
multiplication easily goes through the whole scheme outlined above.

\begin{theorem}\label{th:1}
Let $\fr{g}\sim {\bf B}_r $ or ${\bf D}_r $  and let the dressing factor $
u(x,\lambda ) $ be of the form:
\begin{equation}\label{eq:u-lam}
\fl u(x,\lambda )= \openone + (c_1(\lambda ) -1)P_1(x) + (c_1^{-1}(\lambda )
-1) P_{-1}(x), \qquad P_{-1} =SP_1^TS^{-1},
\end{equation}
where $S $ is introduced in (\ref{eq:br-dr}) and $P_1(x) $ is a rank 1
projector (\ref{eq:pr-1}). Let the constant vectors $|n_0\rangle  $ and
$\langle m_0| $ satisfy the condition
\begin{equation}\label{eq:m-m}
\langle m_0 | S|m_0 \rangle = \langle n_0 | S|n_0 \rangle =0.
\end{equation}
Then $u(x,\lambda ) $ (\ref{eq:u-lam}) satisfies the equation
(\ref{eq:u-eq}) with a potential
\begin{equation}\label{eq:40.6}
q_{(1)}(x) = q_{(0)}(x) - (\lambda _1^+ - \lambda _1^-) [J,p(x)], \qquad
p(x) = P_1(x) - P_{-1}(x).
\end{equation}
\end{theorem}

\begin{proof}
Due to the fact that $\chi_0 ^\pm (x,\lambda ) $ take values in the
corresponding orthogonal group we find that from (\ref{eq:m-m}) it follows
$ \langle m | S|m \rangle =0$, $\langle m | JS|m \rangle =0$ and analogous
relations for the vector $|n\rangle  $.  As a result we get that
\begin{equation}\label{eq:p1p-1}
\fl P_1(x) P_{-1}(x) =P_{-1}(x) P_1(x) =0, \qquad
P_1(x)J P_{-1}(x) =P_{-1}(x)J P_1(x) =0.
\end{equation}

Let us now insert (\ref{eq:u-lam}) into (\ref{eq:u-eq}) and take the
limit of the r.h.side of (\ref{eq:u-eq}) for $\lambda \to \infty $. This
immediately gives eq. (\ref{eq:40.6}). In order that  Eq.~(\ref{eq:u-eq})
be satisfied identically with respect to $\lambda $ we have to put to 0
also the residues of its r.h.side at $\lambda \to \lambda_1^+ $ and
$\lambda \to \lambda_1^- $. This gives us the following system of equation
for the projectors $P_1(x) $ and $P_{-1}(x) $:
\begin{eqnarray}\label{eq:p1-p1}
i {d P_1  \over dx } + q_{(1)}(x) P_1(x) - P_1(x) q_{(0)}(x) - \lambda
_1^- [J, P_1(x)]=0, \\
i {d P_{-1} \over dx } + q_{(1)}(x) P_{-1}(x) - P_{-1}(x) q_{(0)}(x) -
\lambda _1^+ [J, P_{-1}(x)]=0,
\end{eqnarray}
where we have to keep in mind that $q_{(1)} $ is given by (\ref{eq:40.6}).
Taking into account (\ref{eq:p1p-1}) and the relation between $P_1(x) $ and
$P_{-1}(x) $ eq. (\ref{eq:p1-p1}) reduces to:
\begin{equation}\label{eq:P1}
\fl i {d P_1  \over dx } + [q_{(0)}(x), P_1(x)] + \lambda_1^-  P_1(x)J  -
\lambda_1^+ J P_1(x) -(\lambda_1^- -\lambda _1^+) P_1(x) JP_1(x)=0.
\end{equation}
One can check by a direct calculation that (\ref{eq:pr-1}) satisfies
identically (\ref{eq:P1}). The theorem is proved.

\end{proof}

The explicit form of the dressing factor $u(x,\lambda ) $ (\ref{eq:u-lam})
can be viewed as an extension of the results in \cite{VG*87} where
only the structure of the singularities in $\lambda  $ of $u(x,\lambda )
$ or rather of $\chi ^\pm(x,\lambda ) $ was derived for each irreducible
representation of $\fr{g} $.

\begin{corollary}\label{cor:1}
The dressing factor (\ref{eq:u-lam}) can be written in the form
\begin{equation}\label{eq:u-plam}
u(x,\lambda ) = \exp \left( \ln c(\lambda ) \,p(x)\right),
\end{equation}
where $p(x) \in \fr{g} $, and consequently $u(x,\lambda ) $ belongs to the
corresponding orthogonal group.

\end{corollary}

Let us consider now the purely solitonic case, i.e. $q_{(0)}(x)=0 $
and $\chi _{(0)}^\pm(x,t,\lambda )=\exp(-i\lambda (Jx+It)) $.
The condition (\ref{eq:m-m}) which the vector $|n\rangle $ must satisfy
goes into
\begin{eqnarray}\label{eq:nSn}
\sum_{k=1}^{r} 2(-1)^{k} n_{0,k} n_{0,\bar{k}}
= (-1)^{r} \left(n_{0,r+1}\right)^2 ,
\end{eqnarray}
and an analogous one for the vector $|m_0\rangle$. The condition
$\lim_{|x|\to\infty }Q(x,t)=0 $ can be satisfied only if $n_{0k}=0$
whenever $m_{0k}=0 $ and vice versa.
Making use of the explicit form of the projectors $P_{\pm 1}(x) $ valid
for the typical representations of ${\bf B}_r $  we write down:
\begin{equation}\label{eq:p-g}
\fl p(x)= {2  \over \langle m|n\rangle }\left( \sum_{k=1}^{r} h_k(x)
H_{e_k} + \sum_{\alpha \in \Delta _+} ( P_{\alpha }(x) E_{\alpha } +
P_{-\alpha }(x) E_{-\alpha })\right),
\end{equation}
where
\begin{eqnarray}\label{eq:26.1}
\fl h_k(x,t) = n_{0,k}m_{0,k} e^{2\nu _1y_k} - n_{0,\bar{k}}m_{0,\bar{k}}
e^{-2\nu _1y_k} , \nonumber\\
\fl P_{\alpha } = \left\{ \begin{array}{ll}
P_{ks} & \quad \mbox{for\ } \alpha =e_k - e_s, \\
P_{k\bar{s}} & \quad \mbox{for\ } \alpha =e_k + e_s, \\
P_{k,r+1} & \quad \mbox{for\ } \alpha =e_k ,
\end{array} \right. \qquad P_{-\alpha } = \left\{ \begin{array}{ll}
P_{sk} & \quad \mbox{for\ } \alpha =-(e_k - e_s), \\
P_{\bar{s}k} & \quad \mbox{for\ } \alpha =-(e_k + e_s), \\
P_{r+1,k} & \quad \mbox{for\ } \alpha =-e_k ,
\end{array} \right.  \nonumber\\
\fl P_{ks}(x,t) = e^{i\mu _1 (y_s - y_k)} \left( n_{0,k}m_{0,s} e^{\nu
_1(y_k+y_s)} - (-1)^{k+s} n_{0,\bar{s}}m_{0,\bar{k}} e^{-\nu _1(y_k+y_s)}
\right), \\
\mu _1 = {1  \over 2} (\lambda _1^+ + \lambda _1^-), \qquad
\nu _1 = {1  \over 2i} (\lambda _1^+ - \lambda _1^-), \nonumber\\
y_k =J_kx+I_kt , \qquad y_{\bar{k}}=-y_k, \qquad y_{r+1}=0. \nonumber
\end{eqnarray}
Here $1\leq k < s\leq r $ and
\begin{eqnarray}\label{eq:26.5}
\fl \langle m | n \rangle = \sum_{k=1}^{r} \left( n_{0,k}m_{0,k} e^{2\nu
_1y_k} + n_{0,\bar{k}}m_{0,\bar{k}} e^{-2\nu _1y_k} \right) + n_{0,r+1}
m_{0,r+1}.
\end{eqnarray}
The corresponding result for the ${\bf D}_r $-series is obtained formally
if in the above expressions (\ref{eq:26.1}) we put $n_{0,r+1}=m_{0,r+1}=0
$; this means that $P_{k,r+1}=P_{r+1,k}=0 $ for all $k\leq r $. Besides in
the expression for $\langle m|n\rangle  $ (\ref{eq:26.5}) the last term in
the right hand side will be missing.

If we consider the time variable fixed then (\ref{eq:p-g}) and
\begin{equation}\label{eq:26.11}
q_{(1)}(x,t)=-(\lambda _1^+ - \lambda _1^-) \sum_{\alpha \in\Delta _+}
{2 (\alpha ,\vec{J}) \over \langle m|n\rangle }\left(P_\alpha(x) E_\alpha
- P_{-\alpha }(x) E_{-\alpha }\right).
\end{equation}
will provide us the corresponding reflectionless potential of $L $.

Let us note another advantage of the formulae (\ref{eq:p-g}) and
(\ref{eq:26.11}) namely, they provide us $p(x,t) $ and $q(x,t) $ in any
representation of $\fr{g} $.
They also allow us to evaluate explicitly the singularities of both
the FAS and the functions $D_j^+(\lambda ) $ for each of the different
types of solitons. Each type of soliton solution is determined by the set
of nonvanishing components of the vectors $|n_0\rangle $ and $|m_0\rangle
$.  In the general case when all components $n_{0s} $, $m_{0s} $ for
$k_1\leq s\leq k_2 $ are nonzero from (\ref{eq:SD1-0}) we obtain
\begin{equation}\label{eq:30.1}
u_+(\lambda )u^{-1}_-(\lambda ) = \exp \left[ \ln c_1(\lambda )
(H_{\gamma _1} - H_{\gamma _2}) \right].
\end{equation}
By $\gamma _{1,2} $ we denote the weights of the typical representation of
the ${\bf B}_r $-algebra. Thus if we assume that $k_1<r $ and $k_2 \leq r
$ then $\gamma _1 = e_{k_1} $ and $\gamma _2 = e_{k_2} $; for $k_2 =r+1 $
we have $\gamma _{r+1}=0 $ and for $k_2>r+1 $ (therefore $\bar{k}_2 = 2r+2
-k_2 \leq r$) we have $\gamma _{2} =
-e_{\bar{k}_2} $. In order to determine the singularities of the functions
$D_j^+(\lambda ) =\exp (d_j^+(\lambda )) $ we have to use (\ref{eq:5.5}).
The general formula for that is
\begin{equation}\label{eq:30.3}
D_{(1),j}^+(\lambda ) = (c_1(\lambda ))^{b_j} D_{(0),j}^+(\lambda ) ,
\end{equation}
where the integers $b_s = (\gamma _1 - \gamma _2,\omega _s^+) $. Note that
in some cases $\beta =\gamma _1-\gamma _2 $ is not a root. Indeed if
$k_1=k $ and $k_2 =\bar{k} $ then $\beta =2e_k $ and
\begin{equation}\label{eq:30.4}
\fl b_s = \left\{ \begin{array}{ll} 0 &\mbox{for } s<k,\\
2 , & \mbox{for } k \leq s \leq r, \end{array} \right.
\quad \fr{g}\simeq {\bf B}_r,  \qquad
b_s = \left\{ \begin{array}{ll} 0 &\quad \mbox{for } s<k,\\
2 , & \mbox{for } k \leq s \leq r-2, \\
1, & \mbox{for } s= r-1, r. \end{array} \right.
\quad \fr{g}\simeq {\bf D}_r .
\end{equation}

From (\ref{eq:30.4}) we see that in the majority of cases adding a soliton
leads to additional zeroes of the functions $D_j^+(\lambda ) $ of order 2
or 1 in their regions of analyticity; the order depends on the choices of
$k_2 $. There is however one special situation when adding a soliton leads
to additional pole in the region of analyticity. This happens if we have
$\fr{g}\simeq {\bf D}_r $ and $k_1 = r $, $k_2 = \bar{r} $. Then from
(\ref{eq:30.4}) we get
\begin{equation}\label{eq:30.6a}
\fl D_{(1),r-1}^+(\lambda ) = (c_1(\lambda ))^{-1} D_{(0),r-1}^+(\lambda ) ,
\qquad  D_{(1),r}^+(\lambda ) = c_1(\lambda ) D_{(0),r}^+(\lambda ) ,
\end{equation}
i.e., $D_{(1),r-1}^+(\lambda )$ acquires a first-order pole in $\bbbc_+$.

Let us analyze the types of singularities and zeroes of the FAS $\chi
^\pm(x,t,\lambda ) $. Since the ansatz (\ref{eq:u-lam}) for the dressing
factor $u(x,\lambda ) $ contains both positive and negative powers of
$c_1(\lambda ) $ the dressed FAS have both zeroes and poles
in their regions of analyticity. This is one of the important differences
between the analytic properties of the FAS related to the ${\bf A}_r $
and ${\bf B}_r $, ${\bf D}_r $ series. Obviously the order of the poles
and zeroes of FAS will depend on the soliton type and on the
representation of $\fr{g} $ chosen.

The final remark in this section is about the structure of the
different types of soliton solutions. They are determined by the
non-vanishing components of $|n_0\rangle  $ and $|m_0\rangle  $. This in
fact picks up a subset of weights $\Gamma _0 \subset \Gamma (\omega _1) $
in the typical representation of $\fr{g} $. We may relate a subalgebra
$\fr{g}_0 \subset \fr{g} $ whose typical representation is realized on
$\Gamma _0 $. From (\ref{eq:p-g}) we see, that $q_\alpha $ will be
non-zero only if $\alpha \in \Delta _0 $ -- the root system of $\fr{g}_0$.
In Section 4 we give several examples related to different
subalgebras of $\fr{g} $.

\section{Scattering data and the $\bbbz_2 $-reductions.}
\label{5}
Let us address now the question of how the soliton solutions are influenced
by the reductions. To be more specific we consider two important
$\bbbz_2 $ reductions on $U(x,\lambda ) $ (\ref{eq:4.3}), namely:
\begin{eqnarray}\label{eq:34.5}
\mbox{1)} \qquad KU^\dag (x,\lambda ^*) K^{-1} = U(x,\lambda ), \qquad
K^2=\openone ,\\
\label{eq:37.1}
\mbox{2)} \qquad SU(x,-\lambda ) S^{-1} = U(x,\lambda ),
\end{eqnarray}
where $K $ (\ref{eq:35.2}) is an element of the Cartan subgroup of $\fr{g}
$ and $S $ is given by (\ref{eq:br-dr}). Such reductions on $U(x,\lambda )
$ will reflect on $Q(x,t) $ and $J $, and also on the FAS and the
scattering data of $ L(t) $ as follows:
\begin{eqnarray}\label{eq:34.6}
\mbox{1)} \qquad
s_\alpha Q_{-\alpha }^*(x,t)= - Q_\alpha (x,t), \qquad J^*=J, \\
\label{eq:37.3''}
\mbox{2)} \qquad  Q_{-\alpha }(x,t) = Q_\alpha (x,t),
\end{eqnarray}
\begin{eqnarray}\label{eq:ra}
\fl \mbox{1)} &S^+(\lambda ) = K
\left(\hat{S}^-(\lambda ^*)\right)^\dag K^{-1},
\qquad T^+(\lambda ) = K \left(\hat{T}^-(\lambda^*)\right)^\dag K^{-1},
\nonumber\\
&D^+(\lambda ) = K \left(\hat{D}^-(\lambda^*)\right)^\dag K^{-1},
\qquad F(\lambda ) = K \left(F(\lambda ^*)\right)^\dag K^{-1},\\
\label{eq:rc}
\fl \mbox{2)} &S^+(\lambda ) = A_2 \left(S^- (-\lambda)\right)
A_2^{-1}, \qquad
T^+ (\lambda ) = A_2 \left(T^- (-\lambda )\right) A_2^{-1}, \nonumber\\
&D^+ (\lambda ) = A_2 \left(D^- (-\lambda )\right) A_2^{-1},
\qquad F(\lambda ) = A_2 \left(F(-\lambda )\right) A_2^{-1},
\end{eqnarray}
As a minimal set of scattering data we may consider
$T^\pm(\lambda ) $ or $S^\pm(\lambda ) $. The functions $D^\pm(\lambda ) $
or equivalently $d_j^\pm(\lambda)$ can be reconstructed by making use of
their integral representations; in the case of absence of discrete
eigenvalues we have \cite{G*86}:
\begin{equation}\label{eq:5.12}
{\cal  D}_j(\lambda ) = {i  \over 2\pi }\int_{-\infty }^{\infty } {d\mu
\over \mu -\lambda } \ln \langle \omega_j^+|\hat{T}^+(\mu ) T^-(\mu ) |
\omega_j^+\rangle ,
\end{equation}
where $\omega _j^+ $ and $|\omega _j^+\rangle  $ are the $j $-th
fundamental weight of $\fr{g} $ and the highest weight vector in the
corresponding fundamental representation $\Gamma (\omega _j^+) $ of
$\fr{g} $.  The function ${\cal  D}_j(\lambda ) $ as a fraction-analytic
function of $\lambda  $ is equal to:
\begin{equation}\label{eq:5.13}
{\cal  D}_j(\lambda ) = \left\{ \begin{array}{ll}
d_j^+(\lambda ), & \mbox{for} \quad \lambda \in \bbbc_+ \\
(d_j^+(\lambda )-d_{j'}^-(\lambda ))/2, & \mbox{for} \quad \lambda \in
\bbbr, \\
-d_{j'}^-(\lambda ), & \mbox{for} \quad \lambda \in \bbbc_- ,
\end{array} \right.
\end{equation}
where $d_j^\pm (\lambda ) $ were introduced in (\ref{eq:5.5}) and
(\ref{eq:2.13'})  and the index $j' $ is related to $j $ by $w_0(\alpha
_j)=-\alpha _{j'} $. The functions ${\cal  D}_j(\lambda ) $ can be viewed
also as generating functions of the integrals of motion. Indeed, if we
expand
\begin{equation}\label{eq:5.14}
{\cal D}_j(\lambda ) = \sum_{k=1}^{\infty } {\cal D}_{j,k} \lambda^{-k},
\end{equation}
and take into account that $D^\pm(\lambda ) $ are time independent we find
that $d{\cal  D}_{j,k}/dt =0 $ for all $k =1,\dots, \infty $ and
$j=1,\dots r $. Moreover it can be checked that ${\cal  D}_{j,k} $
is local in $Q(x,t) $, i.e. depends only on $Q $ and its derivatives with
respect to $x $.

{}From (\ref{eq:5.12}) and (\ref{eq:ra})-(\ref{eq:rc}) we easily obtain the
effect of the reductions on the set of integrals of motion; namely, for
the reduction (\ref{eq:ra}):
\begin{equation}\label{eq:5.15a}
{\cal  D}_j(\lambda )=- {\cal  D}_j^*(\lambda ^*), \qquad
\mbox{i.e.}, \qquad {\cal  D}_{j,k} = - {\cal  D}_{j,k}^*,
\end{equation}
and for (\ref{eq:rc})
\begin{equation}\label{eq:5.15b}
{\cal  D}_j(\lambda )=-{\cal  D}_j(-\lambda ), \qquad
\mbox{i.e.}, \qquad {\cal  D}_{j,k} = (-1) ^{k+1} {\cal  D}_{j,k}.
\end{equation}

From (\ref{eq:5.15b}) it follows that al integrals of motion
with even $k $ become degenerate. The
reduction (\ref{eq:5.15a}) means that the integrals ${\cal  D}_{j,k} $
are purely imaginary; if we impose in addition also (\ref{eq:5.15b}) then
${\cal D}_{j,2k}=0 $.
The simplest local integrals of motion ${\cal D}_{j,1} $ and ${\cal
D}_{j,2} $ can be expressed as functionals of the potential $Q $ of
(\ref{eq:1.1}) as follows (see \cite{G*86}):
\begin{eqnarray}\label{eq:Dj1}
&&
{\cal D}_{j,1} = -{ i \over 4 } \int_{-\infty }^{\infty } dx\, \langle [J,
Q], [H_j^{\vee} ,Q] \rangle ,
\end{eqnarray}
\begin{eqnarray}\label{eq:Dj2}
\fl &&
{\cal D}_{j,2} = -{ 1\over 2 } \int_{-\infty }^{\infty } dx\, \langle
Q, [H_j^{\vee} ,Q_x] \rangle  -{ i \over 3 } \int_{-\infty }^{\infty }
dx\, \langle [J, Q], [Q, [H_j^{\vee} ,Q]] \rangle ,
\end{eqnarray}
where $\langle H_j^{\vee}, H_{k}\rangle =\delta _{jk} $.
The fact that ${\cal D}_{j,1} $ are integrals of motion for $j=1,\dots, r
$, can be considered as natural analog of the Manley--Rowe relations
\cite{ZM,K}. In the case when the reduction is of the type (\ref{eq:C-1}),
i.e. $Q_{-\alpha} =s_\alpha Q_\alpha ^* $ then (\ref{eq:Dj1}) is equivalent
to
\begin{eqnarray}\label{eq:M-R}
\sum_{\alpha >0} {2 (\vec{J},\alpha )(\omega _j^+,\alpha ) \over (\alpha
,\alpha )}\int_{-\infty }^{\infty }dx\, s_{\alpha } |Q_{\alpha }(x)|^2 =
\mbox{const} ,
\end{eqnarray}
and can be interpreted as relations between the wave densities $|Q_\alpha
|^2 $.

The Hamiltonian of the $N $-wave equations (\ref{eq:1.4}) is expressed
through ${\cal D}_{j,2} $, namely:
\begin{eqnarray}\label{eq:H-D}
H_{N\rm -wave} &=& -\sum_{j=1}^{r}
{2(\alpha _j, \vec{I})\over (\alpha _j, \alpha _j)}  {\cal D}_{j,2} =
{1 \over 2i} \left\langle  \left\langle \dot{{\cal  D}}(\lambda ),
{}f(\lambda ) \right\rangle \right\rangle _{0},
\end{eqnarray}
where $\dot{{\cal  D}}(\lambda )=d{\cal  D}/d\lambda $ and
$f(\lambda )=\lambda I $ is the dispersion law of the $N $-wave equation
(\ref{eq:1.4}). In (\ref{eq:H-D}) we used just one of the hierarchy of
scalar products in the Kac-Moody algebra $\widehat{\fr{g}} \equiv \fr{g}
\otimes \bbbc[\lambda ,\lambda ^{-1}] $ (see \cite{RST}):
\begin{equation}\label{eq:ScPr}
\fl \left\langle  \left\langle X(\lambda ), Y(\lambda ) \right\rangle
\right\rangle _{k} = \mbox{Res} \, \lambda^{k+1} \left\langle
\hat{D}^+(\lambda ) X(\lambda ),Y(\lambda )\right\rangle , \qquad
X(\lambda ), Y(\lambda ) \in \widehat{\fr{g}}.
\end{equation}

\section{Soliton solutions and examples of $N $-wave equations.}

\subsection{Examples of soliton solutions}

Here we will give several examples of $1 $-soliton solutions of the $N
$-wave equations subject to the reductions (\ref{eq:34.5}) and
(\ref{eq:37.1}). Obviously the dressing factor $u(x,\lambda ) $ must
satisfy the same reduction conditions as $U(x,\lambda ) $. This allows us
to derive the following consequences on the  vectors $|n_0\rangle  $,
$|m_0\rangle  $ and $\lambda _1^\pm $:
\begin{eqnarray}\label{eq:34.8}
\mbox{1)} \qquad |n_0\rangle =K |m_0^*\rangle , \qquad \lambda_1^-
=(\lambda _1^+)^*, \\
\label{eq:37.8}
\mbox{2)} \qquad |n_0\rangle = |m_0\rangle , \qquad \lambda_1^- =-\lambda
_1^+,
\end{eqnarray}

\begin{example}\label{exa:1}
Let $\fr{g}_0\simeq sl(2) $ be formed by the generators $H_\alpha  $ and
$E_{\pm\alpha } $ with $\alpha =e_k-e_s $. This is possible if only two of
the components $n_{0k} $, $n_{0s} $ are non-vanishing with $1\leq k <s
\leq r $. Then:
\begin{eqnarray}\label{eq:75.10}
\fl p(x)= {1 \over 2} H_{e_k+e_s} + {1 \over 2 \cosh \Phi} \left( \sinh
\Phi H_\alpha + e^{-i\Psi } E_{\alpha } + e^{i\Psi } E_{-\alpha }\right),
\\
\fl \Phi(x,t) = \nu _1(y_k - y_s) + {1  \over 2 } \ln { n_{0,k}
m_{0,k}\over n_{0,s} m_{0,s} } , \qquad \Psi(x,t) =\mu _1(y _k -y_s) + {i
\over 2 } \ln {n_{0,k} m_{0,s} \over n_{0,s} m_{0,k} } ,\nonumber
\end{eqnarray}
If we apply the reduction 1) we find that $\Phi  $ and $\Psi  $ become
real provided $K_s=K_k $; the corresponding soliton solution is regular
exponentially decaying function for all values of $t $. If $K_kK_s=-1 $
then $\Phi =\nu _1(y_k-y_s) + \ln (|n_{0,k}/n_{0,s}| -i\pi/2 $ which
effectively changes the denominators in (\ref{eq:75.10}) to $\sinh \Phi$
thus making the soliton solution singular. Such solution require
additional care.

Imposing reduction 2) the arguments of $p(x,t) $ simplify to
\[
\Phi (x,t) = \nu _1(y_k-y_s) + \ln {n_{0k}  \over n_{0,s} }, \qquad
\Psi (x,t) = \mu _1 (y_k-y_s).
\]
If we apply both reduction simultaneously then the eigenvalues of $L $
become purely imaginary $\lambda _1^+=-\lambda _1^-=i\nu _1 $, $\Phi (x,t)
$ remains as in the last line and $\Psi (x,t) $ vanishes since $\mu _1=0 $.

We can evaluate also the solitons related to another choice of $\alpha
=e_k+e_s $; then we must assume that the non-vanishing  $n_{0k} $,
$n_{0s} $ are those with $1\leq k <\bar{s} \leq r $. The explicit formulae
are analogous to the ones above and we skip them.

\end{example}

\begin{example}\label{exa:2}
The solitons related to the subalgebra $\fr{g}_0\simeq so(3) $ spanned by
$H_{\alpha } $ and $E_{\pm\alpha } $ with $\alpha =e_k $ are obtained
by choosing $n_{0k} $, $n_{0,r+1} $ and $n_{0,\bar{k}} $, $k\leq r $ as
the only non-vanishing components. Due to (\ref{eq:nSn}) we
must have $(n_{0,r+1})^2=2n_{0,k}n_{0,\bar{k}} $ and:
\begin{eqnarray}\label{eq:79.1}
\fl  p(x)=  \tanh \Phi  H_\alpha + {e^{\Phi } + (-1)^{k+r} e^{-\Phi }
\over 2\sqrt{2} \cosh^2 \Phi  } \left( e^{-i\Psi } E_\alpha + e^{i\Psi }
E_{-\alpha }\right), \\
\fl \Phi(x,t) = \nu _1y_k + {1  \over 4 } \ln {n_{0,k}m_{0,k}  \over
n_{0,\bar{k}}m_{0,\bar{k}} }, \qquad  \Psi (x,t) = \mu _1y_k +
{1  \over 4 } \ln {n_{0,k}m_{0,\bar{k}} \over n_{0,\bar{k}}m_{0,k} }.
\nonumber
\end{eqnarray}
These solitons are not singular due to the fact that $K_k = K_{\bar{k}} $.
The reduction 1) makes both $\Phi (x,t) $  and $\Psi (x,t) $ real.
Under the reduction 2) $\Psi  $ vanishes as in the first example.

\end{example}

\begin{example}\label{exa:3}
The solitons related to the subalgebra $\fr{g}_0\simeq sl(3) $ whose
positive roots are $e_i-e_k $, $e_k-e_j $ and $e_i-e_j $
are obtained if we choose as the only non-vanishing components $n_{0,i} $,
$n_{0,k} $ and $n_{0,j} $, $i<k< j\leq r $; the condition (\ref{eq:nSn})
holds identically. Here we directly write down the formulae with reduction
1) applied.
\begin{eqnarray}\label{eq:84.3}
\fl p(x)= {1 \over 3}  H_{\vec{\varepsilon }} + {1  \over 3\langle n^*|
K|n\rangle }\left( (2h_i - h_j - h_k)(x,t) H_{e_i-e_k} +
(2h_j -h_k -h_i)(x,t) H_{e_j-e_k}\right) \nonumber\\
+ {1  \over \langle n^*| K|n\rangle  } \left( N_{ik}(x,t) +
N_{kj}(x,t) + N_{ij}(x,t)\right),\\
\langle n^*| K|n\rangle = (h_i + h_j + h_k)(x,t), \qquad
h_i(x,t) = K_i e^{\Phi _{i}}, \nonumber\\
N_{ik}(x,t)= e^{\Phi _{ik}} \left( e^{-i\Psi _{ik}} E_{e_i-e_k} +
e^{i\Psi _{ik}}  E_{-e_i+e_k} \right), \qquad
\vec{\varepsilon }=e_i + e_j + e_k, \nonumber \\
\fl \Phi _{ik}(x,t) = \nu _1(y_i + y_k) + \ln |n_{0,i} n_{0,k}|, \quad
\Psi _{ik}(x,t) = \mu _1 (y_i-y_k) +\arg n_{0,k} - \arg n_{0,i},
\nonumber
\end{eqnarray}
This result for $K_k=K_i=K_j $ is quite analogous to the regular soliton
solutions of the $3$- and $N $-wave equations studied in detail in
\cite{1}. The situation when $K_k=-K_i=-K_j $ again may lead to singular
solitons which describe blow-up instabilities in $\chi ^{(2)} $-media.
\end{example}

In all the examples above the corresponding reflectionless potentials of $
L $ are obtained with the formula (\ref{eq:26.11}).

\subsection{Real forms of ${\bf B}_2 $.}

Let us illustrate these general results by a an example related to the
${\bf B}_2$ algebra. This algebra has two simple roots $\alpha_1=e_1-e_2$,
$\alpha _2=e_2 $, and two more positive roots: $\alpha _1+\alpha _2=e_1
$ and $ \alpha _1+2\alpha _2= e_1+e_2=\alpha _{\rm max} $. When they come
as indices, e.g. in $Q_\alpha  $ we will replace them by sequences of two
integers: $\alpha \to kn $ if $\alpha =k\alpha _1 + n\alpha _2 $; if
$\alpha =-(k\alpha _1 + n\alpha _2) $ we will use $\overline{kn} $.

The reduction which extracts the real forms of ${\bf B}_2\simeq so(5) $
is  $KU^{\dag}(\lambda^*)K^{-1} =U(\lambda )$ where $K $ is an element of
the Cartan subgroup: $K=\mbox{diag} \, (s_1, s_2, 1, s_2, s_1) $ with
$s_k=\pm 1 $, $k=1,2 $.  This means that $J_i=J_i^* $, $i=1,2 $ and
$Q_\alpha  $ must satisfy:
\begin{eqnarray}\label{eq:b2.1}
\fl p_{10}=-s_2 s_1 Q_{10}^*, \quad p_{01}=-s_2 Q_{01}^*, \quad p_{11}=
- s_1 Q_{11}^*, \quad p_{12}=- s_1s_2 Q_{12}^*.
\end{eqnarray}
Thus we get 4-wave system with the Hamiltonian $H=H_0 +H_{\rm int} $
\cite{3}:
\begin{eqnarray}\label{eq:b2.2}
\fl &&H_0=-{i\over 2}\int_{-\infty }^{\infty }dx \, \left[s_1s_2(I_1-I_2)
(Q_{10}Q_{10,x}^*-Q_{10,x}Q_{10}^*) +
2s_2I_2(Q_{01}Q_{01,x}^*-Q_{01,x}Q_{01}^*)  \right. \nonumber\\
\fl &&\qquad +\left. 2 s_1I_1(Q_{11}Q_{11,x}^*-Q_{11,x}Q_{11}^*) +
s_1s_2(I_1+I_2) (Q_{12}Q_{12,x}^*- Q_{12,x}Q_{12}^*) \right], \\
\fl &&H_{{\rm int}} = 2\kappa s_1\int_{-\infty }^{\infty }dx\,
\left[s_2 (Q_{12}Q_{11}^*Q_{01}^*+Q_{12}^*Q_{11}Q_{01}) +
(Q_{11}Q_{01}^*Q_{10}^*+Q_{11}^*Q_{01}Q_{10})\right], \nonumber
\end{eqnarray}
where $\kappa =J_1I_2-J_2I_1 $, and the symplectic 2-form:
\begin{eqnarray}\label{eq:b2.2a}
\Omega ^{(0)} &=& i \int_{-\infty }^{\infty } dx \, [ (J_1-J_2)\delta
Q_{10} \wedge \delta Q_{10}^* + 2J_2\delta Q_{01} \wedge \delta Q_{01}^*
\nonumber\\
&+&2J_1\delta Q_{11} \wedge \delta Q_{11}^*+ (J_1+J_2)\delta Q_{12} \wedge
\delta Q_{12}^*],
\end{eqnarray}
The corresponding wave-decay diagram is shown in figure 1.

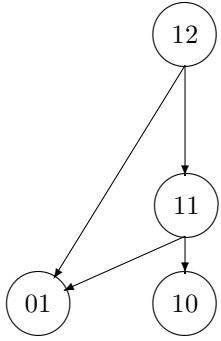
\begin{figure}{}
\caption{\label{fig:1}
Wave-decay diagram for the $so(5) $ algebra. To each positive
root of the algebra $\underline{\rm kn}\equiv k\alpha _1 + n\alpha _2 $
we put in correspondence a wave of type $\underline{\rm kn} $. If the
positive root $\underline{\rm kn} = \underline{\rm k'n'}+\underline{\rm
k''n''} $ can be represented as a sum of two other positive roots, we say
that the wave $\underline{\rm kn} $ decays into the waves $\underline{\rm
k'n'} $ and $\underline{\rm k''n''} $ as shown on the diagram to the left.}
\vspace{0.3cm}
\special{em:linewidth 0.4pt}
\unitlength 0.65mm
\linethickness{0.4pt}
\begin{picture}(66.33,81.00)
\put(30.00,20.00){\circle{12.00}}
\put(30.00,20.00){\makebox(0,0)[cc]{01}}
\put(60.00,20.00){\circle{12.00}}
\put(60.00,20.00){\makebox(0,0)[cc]{10}}
\put(60.33,40.00){\circle{12.00}}
\put(60.33,40.00){\makebox(0,0)[cc]{11}}
\put(60.00,75.00){\circle{12.00}}
\put(60.00,75.00){\makebox(0,0)[cc]{12}}
\put(33.33,25.33){\vector(-2,-3){0.2}}
\emline{60.00}{68.67}{1}{33.33}{25.33}{2}
\put(60.00,46.33){\vector(0,-1){0.2}}
\emline{60.00}{68.67}{3}{60.00}{46.33}{4}
\put(60.00,26.33){\vector(0,-1){0.2}}
\emline{60.00}{33.67}{5}{60.00}{26.33}{6}
\put(35.33,22.67){\vector(-2,-1){0.2}}
\emline{60.00}{33.67}{7}{35.33}{22.67}{8}
\end{picture}
\hspace*{-4cm}
\end{figure}

The particular case $s_1=s_2=1 $ leads to the compact real form
$so(5,0)\simeq so(5,\bbbr) $ of the ${\bf B}_2 $-algebra. The choice
$s_1=-s_2=-1 $ leads to the noncompact real form $so(2,3) $ and
$s_1=s_2=-1 $ gives another noncompact one-- $so(1,4) $. If in this last
case we identify $Q_{01}=Q_{\rm pol} $, $Q_{10}=-E_{\rm s} $,
$Q_{11}=E_{\rm p} $ and $Q_{12}=-E_{\rm a} $ then we obtain the system
studied in \cite{3} which describes Stockes-anti-Stockes wave generation.
Here $Q_{\rm pol} $ is the normalized effective polarization of the medium
and $E_{\rm p} $, $E_{\rm s} $ and $E_{\rm a} $ are the normalized pump,
Stockes and anti-Stockes wave amplitudes respectively.

\subsection{One more $\bbbz _2$ reduction}

Let us apply a second $\bbbz_2$-reduction to the already reduced
system of the previous subsection. We take it of the form
(\ref{eq:37.1}) which gives $J_i=J_i^* $, $I_i=I_i^* $
and:
\begin{eqnarray}\label{eq:b2.3}
\fl Q_{10}^*=-s_1 s_2 Q_{10}, \quad Q_{01}^*=-s_2Q_{01},
\quad Q_{11}^*=-s_1Q_{11}, \quad Q_{12}^*=-s_1s_2Q_{12}.
\end{eqnarray}
These reduction conditions allow us to make the following change of the
fields $Q_{\alpha } $ to the real-valued ones $v_{\alpha } $ as follows:
\begin{eqnarray}\label{eq:q-v}
\fl Q_{10}=i^{(1+s_1s_2)/2}v_{10}, \quad Q_{01}=i^{(1+s_2)/2}v_{01}, \quad
Q_{11}=i^{(1+s_1)/2}v_{11}, \quad Q_{12}=i^{(1+s_1s_2)/2}v_{12}.
\nonumber
\end{eqnarray}
Thus we get the following 4-wave system for 4 real-valued
functions:
\begin{eqnarray}\label{eq:b2.4}
&&(J_1-J_2)v_{10,t}-(I_1-I_2)v_{10,x}+2s_2\kappa v_{11}v_{01}=0,
\nonumber\\
&&J_2v_{01,t}-I_2v_{01,x}+s_1s_2\kappa (v_{11}v_{12}+v_{11}v_{10})=0,
\nonumber\\
&&J_1v_{11,t}-I_1v_{11,x}+\kappa (v_{12}v_{01}-v_{10}v_{01})=0, \\
&&(J_1+J_2)v_{12,t}-(I_1+I_2)v_{12,x}-2s_2\kappa v_{11}v_{01}=0.
\nonumber
\end{eqnarray}
Since $w_0(J)=-J $ the Hamiltonian structure $\{H^{(0)}, \Omega
^{(0)}\} $ becomes degenerated.

Such reductions applied to the Zakharov-Shabat system with $\fr{g}\simeq
sl(2) $ picks up the sine-Gordon and the MKdV equations which have two
types of soliton solutions: i)~ones related to pairs $\pm i\nu _k $ of
purely imaginary eigenvalues of $L $ and ii)~ones related to quadruplets
$\pm \lambda _j^+$, $ \pm \lambda _j^* $ of eigenvalues of $L $. The first
type are known as `topological' solitons, while the second type are known
as `breathers'.

The same situation persists also for $L $ (\ref{eq:1.4}) related to any
$\fr{g} $. One of the differences between $\fr{g}\simeq sl(2) $ and
generic $\fr{g} $ consists also in the fact that in the $sl(2) $
case the $N $-wave interaction is trivial, while for higher ranks of
$\fr{g} $ it is non-trivial, see e.g.  (\ref{eq:b2.4}).

\section{Hamiltonian structures of the reduced $N $-wave
equations}\label{sec:6}

The generic $N $-wave interactions (i.e., prior to any reductions) possess
a hierarchy of Hamiltonian  structures generated by the so-called
generating (or recursion) operator $\Lambda $ defined in \cite{G*86,VG*86}
as follows:
\begin{eqnarray}\label{eq:hs-k}
\Omega^{(k)} = {i \over 2} \int_{-\infty }^{\infty } dx\, \left\langle
[J, \delta Q(x,t)] \wedgecomma \Lambda ^k \delta Q(x,t) \right\rangle ,
\end{eqnarray}
Using the spectral decomposition of $\Lambda  $ we can recalculate $\Omega
^{(k)} $ in terms of the scattering data of $L $ with the result:
\begin{eqnarray}\label{eq:5.17}
&& \Omega ^{(k)} = {1 \over 2\pi } \int_{-\infty }^{\infty } d\lambda
\lambda ^k \left( \Omega _0^+(\lambda ) - \Omega _0^-(\lambda )\right),
\nonumber\\
&& \Omega _0^\pm(\lambda ) = \left\langle \hat{D}^\pm(\lambda )
\hat{T}^\mp(\lambda ) \delta T^\mp(\lambda ) D^\pm(\lambda ) \wedgecomma
\hat{S}^\pm(\lambda ) \delta S^\pm(\lambda ) \right\rangle .
\end{eqnarray}
The first consequence from (\ref{eq:5.17}) is that the kernels of
$\Omega^{(k)} $ differs only by the factor $\lambda ^k $; i.e., all of
them can be cast into canonical form simultaneously. This is quite
compatible with the results of \cite{ZM,ZM1,BS} for the action-angle
variables.

Again it is not difficult to find how the reductions influence $\Omega
^{(k)} $. Using the invariance of the Killing form, from (\ref{eq:5.17})
and (\ref{eq:ra})--(\ref{eq:rc}) we get respectively: $\Omega ^+_0(\lambda
) = (\Omega _0^-( \lambda ^*))^*$, and $\Omega ^+_0(\lambda ) = \Omega
_0^-(-\lambda)$.
Therefore for the reduction (\ref{eq:ra}) we get that  $i\Omega ^{(k)} $
together with all the integrals $i{\cal  D}_{j,k} $ become simultaneously
real. The other reduction (\ref{eq:rc}) means that $\Omega ^{(k)} =
(-1)^{k+1}\Omega ^{(k)} $; besides from (\ref{eq:5.15b})  we have ${\cal
D}_{j,2k}=0 $. Thus `half' of the Hamiltonian structures  $\left\{ \Omega
^{(p)}, H^{(p)} \right\}$ with even $p $ degenerate.  However the other
`half' for odd $p $ survives. In particular this means that the canonical
2-form $\Omega ^{(0)} $ is also degenerate, so the $N $-wave equations
with the reduction (\ref{eq:5.15b}) do not allow Hamiltonian formulation
with canonical Poisson brackets. For more details see \cite{2,VG*86,VYa}.

\section{Conclusions}\label{sec:5}

We end with several remarks.

{\bf 1.} Here we presented only examples related to the ${\bf B}_2 $
algebra. Many additional examples can be found in \cite{GGK}.

{\bf 2.} To all reduced systems given above we can apply the analysis in
\cite{G*86,VYa} and derive the spectral decompositions for the
corresponding recursion $\Lambda  $ operator. Such analysis allows  one to
prove the pair-wise compatibility of the Hamiltonian structures.

{\bf 3.} In many cases the reduction conditions on $J $ may lead to lead
to complex values of $J_k $. The construction of the corresponding FAS is
first given in \cite{BC} for $sl(n) $ and in \cite{VYa} for simple
$\fr{g}$. It is an interesting task to extend the results of \cite{BC,VYa}
to systems with reductions.

{\bf 4.} The cases when the element $J $ is not regular requires
additional care in constructing the theory of the recursion operator
$\Lambda  $, see \cite{VSG*94}.

{\bf 5.} The list of examples with soliton solutions can easily be
continued.  Several factors are important for the structure of the
different types of solitons. These are:  the rank of the projectors $P_{1}
$ and $P_{-1} $ and the subalgebra $\fr{g}_0 $ which is picked up.
Obviously depending on these choices we get solitons with different number
of internal degrees of freedom. These questions will be discussed
elsewhere.
Some alternative approaches for constructing soliton solutions of
reduced systems are presented in \cite{Cho*99,TerUl}.

\section*{Acknowledgements} We thank two of the referees for useful
suggestions.


\section*{References}
\begin{thebibliography}{999}

\bibitem{ZM}
Zakharov V E, Manakov S V.,
{\it Exact theory of resonant interaction of wave packets in nonlinear
media}, INF preprint 74-41, Novosibirsk (1975); (In Russian);

\bibitem{ZM1}
Zakharov V E, Manakov S V, Zh. Exp. Teor. Fiz, {\bf
69}, 1654--1673 (1975);  (In Russian); \\
Manakov S V, Zakharov V E,  Zh.  Exp.
Teor.  Fiz.  {\bf 71}, 203--215 (1976) (In Russian).

\bibitem{K} Kaup D J.,
Stud. Appl. Math, {\bf 55}, 9--44 (1976).

\bibitem{1} Zakharov V E,  Manakov S V, Novikov S P.,
Pitaevskii L I, {\it Theory of solitons: the inverse scattering method},
Plenum, N.Y, 1984;

\bibitem{FaTa} Faddeev L D, Takhtadjan L A, {\it Hamiltonian approach
in the theory of solitons\/}, Springer Verlag, Berlin, (1987).

\bibitem{KRB} Kaup D J, Reiman A, Bers A.,
Rev.  Mod.  Phys., {\bf 51}, 275--310 (1979).

\bibitem{Sh} Shabat A B, Functional Annal.  \& Appl.  {\bf 9}, n.3,
75--78 (1975) (In Russian);\\
Shabat A B,  Diff.  Equations {\bf 15}, 1824--1834 (1979) (In Russian).

\bibitem{Za*Sh} Zakharov V E, Shabat A B, Funkt. Anal. i
Pril.  {\bf 8} (1974) 43-53; (In Russian) \\
Zakharov V E, Shabat A B,  Funkt. Anal. i Pril, {\bf 13} (1979),
13-22 (In Russian).

\bibitem{G*86}   Gerdjikov V S, Kulish P P.,
   Physica D, {\bf 3}, n 3, 549--564, 1981.

\bibitem{BS}
Beals R, Sattinger D.,
Commun. Math.  Phys.  {\bf 138} no 3, 409--436 (1991).

\bibitem{VG*87}   Gerdjikov V S.,
Phys.  Lett.  A {\bf 126}, n 3, 184--186, 1987.

\bibitem{Za*Mi}  Zakharov V E,  Mikhailov A V,
Commun. Math. Phys. {\bf 74} (1980) 21--40;\\
   Zakharov~V~E, Mikhailov~A~V.
    {\it Zh.~Eksp. Teor. Fiz.} {\bf 74} 1953, (1978).

\bibitem{VG*86}   Gerdjikov V S, Inverse Problems {\bf 2}, n 1, 51--74,
1986.


\bibitem{LMP}   Gerdjikov V.~S.
   Lett.\ Math.\ Phys.\ {\bf 6}, n.~6, 315--324, 1982.

\bibitem{2} Mikhailov A V.,
Physica D, {\bf 3D}, n 1/2, 73--117 (1981).

\bibitem{MiOlPer} Mikhailov A V,  Olshanetzky M A,  Perelomov A M.
   Commun.\ Math.\ Phys.\ {\bf 79}, 473--490, 1981.

\bibitem{AKNS}
   Ablowitz M J, Kaup D J, Newell A C, Segur H.
   Studies in Appl.\ Math.\ {\bf 53}, n 4, 249--315, 1974.

\bibitem{ForGib}   Fordy A P,  Gibbons J.,
   Commun.\ Math.\ Phys.\ {\bf 77}, n 1, 21--30 (1980).

\bibitem{ForKu}
  {} Fordy A P,  Kulish P P.,
   Commun.\ Math.\ Phys.\ {\bf 89}, n 4, 427--443 (1983).

\bibitem{LA}  Bourbaki N.,
{\it Elements de mathematique. Groupes et algebres de Lie. Chapters
I--VIII},  Hermann, Paris (1960--1975).\\
Goto M, Grosshans F, {\it Semisimple Lie algebras}, Lecture Notes in
Pure and Applied Mathematics vol. {\bf 38}, M.Dekker Inc, New York \&
Basel 1978.

\bibitem{Helg}  Helgasson S.,
{\it Differential geometry, Lie groups and symmetric spaces},
   Academic Press, 1978.

\bibitem{RST} Kulish. P. P, Reyman A G,  Zap. Nauch. Semin. LOMI, {\bf
123} (1983),67--76 (In Russian); \\
Reyman A. G,  Zap.  Nauch.  Semin.  LOMI, {\bf 131}
(1983),118--127 (In Russian).

\bibitem{3} Gerdjikov V S,   Kostov N A,
Phys Rev {\bf 54 A}, 4339--4350 (1996);\\ Gerdjikov V S,  Kostov
N A,   {\bf patt-sol/9502001.}

\bibitem{VYa}
Gerdjikov V S, Yanovski A B,  J. Math. Phys. {\bf 35},
no 7, 3687--3725 (1994).

\bibitem{GGK}
Gerdjikov V S,  Grahovski G G, Kostov N A.
In: Eds. Boiti M,  Martina L, Pempinelli F, Prinari B, Soliani G.
``Nonlinearity, integrability and all that: 20
years after NEEDS~79'', pp. 279-283, World Scientific,  1999;\\
\dash In: Eds. Ivailo M. Mladenov and Gregory L. Naber.,
"Geometry, Integrability and Quantization",
Coral Press Scientific Publications, pp. 55--77, Sofia, (2000);\\
\dash {\bf nlin.SI/0006001}.

\bibitem{BC}
   Beals R, Coifman R R, Commun.\ Pure and Appl.\ Math.\ {\bf 37},
n 1, 39--90, 1984.

\bibitem{VSG*94} Gerdjikov V G,  Teor. Mat. Fiz. {\bf 99} (1994),
292-299.


\bibitem{Cho*99} Choudhury A G,  Choudhury A R,
Int. J. Mod. Phys. {\bf A14}, 3871-3883 (1999).

\bibitem{TerUl} C-L. Terng, K. Uhlenbeck. Commun. Pure \& Appl. Math. {\bf
53}, 1--75 (2000).

\end{thebibliography}
\end{document}